\documentclass[aps,prl,reprint,superscriptaddress,longbibliography,amsmath,amssymb]{revtex4-1}

\usepackage[final]{graphicx}
\usepackage{hyperref}
\usepackage{color}

\DeclareMathOperator{\Tr}{Tr}
\DeclareMathOperator{\Real}{Re}

\begin{document}


\title{Stroboscopic Tests for Thermalization of Electrons in Pump/Probe Experiments}

\author{O.~P.~Matveev}
\affiliation{Institute for Condensed Matter Physics of the National Academy of Sciences of Ukraine, Lviv, 79011 Ukraine}
\author{A.~M.~Shvaika}
\affiliation{Institute for Condensed Matter Physics of the National Academy of Sciences of Ukraine, Lviv, 79011 Ukraine}
\author{T.~P.~Devereaux}
\affiliation{Geballe Laboratory for Advanced Materials, Stanford University, Stanford, CA 94305,USA}
\affiliation{Stanford Institute for Materials and Energy Sciences (SIMES), SLAC National Accelerator Laboratory, Menlo Park, CA 94025, USA}
\author{J.~K.~Freericks}
\affiliation{Department of Physics, Georgetown University, Washington, DC 20057, USA}

\begin{abstract}
One of the goals of pump/probe spectroscopies is to determine how electrons relax after they have
been driven out of equilibrium. It is challenging to determine how close electrons are to a thermal
state solely by fitting their distribution to a Fermi-Dirac distribution. Instead, we propose that 
one compare the effective temperatures of both fermions and collective bosonic modes (derived
from the fermions) to determine the distance from a thermal state. Measurements of effective fermionic and bosonic temperatures can be achieved directly via photoemission and nonresonant Raman scattering. Their difference quantifies the distance from thermal equilibrium.              
\end{abstract}


\maketitle

Time-resolved pump/probe experiments are commonly used to examine the nonequilibrium dynamics of different excitations in solids~\cite{hsieh:035128,hellmann:187401,perfetti:067402,wegkamp:216401,smallwood:27001,giannetti:58}. 
Two of the most common spectroscopies measured in these experiments are optical conductivity
and photoemission spectroscopy. Here we focus on how another spectroscopy---nonresonant electronic Raman scattering---can be combined with photoemission to determine proximity of the electrons to thermal equilibrium.  It is well known that thermometry of electrons is challenging
in ultrafast experiments and often is performed by fitting to a Fermi-Dirac distribution function. Experimentally, nonequilibrium Raman scattering has been measured for phonons in graphite~\cite{yang:40876} and combined with photoemission of electrons to study the thermal relaxation of both.   Here, we develop the theory for nonresonant electronic Raman scattering  in the $B_{\text{1g}}$ symmetry channel and combine it with the known methods for photoemission to compare the two different effective temperatures.
We perform our calculations for the spinless Falicov-Kimball model~\cite{falicov:997} within the nonequilibrium extension~\cite{freericks:266408} of dynamical mean-field theory (DMFT)~\cite{brandt:365,freericks:1333}. The model describes a strongly correlated metal-Mott-insulator transition, which occurs at the critical Coulomb interaction $U_c=\sqrt{2}$ when the system is at half filling. 

In a pump-probe setup with a nonequilibrium pump and finite width probes, we adapt the formalism proposed by Nozi\`{e}res and Abrahams~\cite{nozieres:3099} for resonant inelastic X-ray scattering (RIXS) to nonresonant electronic Raman scattering.
The quantum system evolves from $t_1$ to $t_2$ via the evolution operator $ U(t_2, t_1)=\mathcal{T}_t\exp{\bigl\{-i\int^{t_2}_{t_1}d\tilde{t}\mathcal{H}(\tilde{t})\bigr\}}$. Initially ($t\to-\infty$), the system is in an equilibrium state $|n\rangle$, which is an eigenstate of the initial (electronic) Hamiltonian $\mathcal H_0=\mathcal H(t\to-\infty)$. The system has two electric fields applied to it. The pump, which is treated semiclassically and the probe, which is treated quantum mechanically and perturbatively. 
The full system at arbitrary time $t$ is then   
\begin{align}\label{evolution}
& |\psi(t)\rangle=U(t,-\infty)|n\rangle\otimes a_{\mathbf{k}_i, \mathbf{e}_i}^\dagger|0\rangle
\nonumber \\
 & =\mathcal{T}_t\exp{\biggl\{-i\int^{t}_{-\infty}d\tilde{t}(\mathcal{H}_{\text{loc}}+\mathcal{H}_{t}(\tilde{t}))\biggr\}}|n\rangle\otimes a_{\mathbf{k}_i, \mathbf{e}_i}^\dagger|0\rangle,
\end{align}
where $a_{\mathbf{k}_i, \mathbf{e}_i}^{\dag}$ creates an incident photon with momentum ${\mathbf{k}_i}$, energy $\omega_i$, and polarization $\mathbf{e}_i$. The electronic Hamiltonian has two parts---$\mathcal{H}_{\text{loc}}=\sum_i [U n_{ic} n_{if} - \mu n_{ic} + E_f n_{if}]$ includes the local interaction $U$ between itinerant ($c$) and localized ($f$) electrons (and their chemical potentials) and the time-dependent part of the Hamiltonian $\mathcal{H}_{t}(t)$ describes the interaction with the total electric field (via the Peierls' substitution to the hopping term):  
\begin{equation}\label{int}
 \mathcal{H}_{t}(t)=-\sum\limits_{\langle i,j\rangle}\frac{t^{*}}{2\sqrt{D}}e^{-i\int^{\mathbf{R}_{j}}_{\mathbf{R}_{i}}d\mathbf{r} \cdot\mathbf{A}(r,t)}c^{\dag}_{i}c_{j}^{\phantom\dagger}.
\end{equation}
Here $n_{ic}=c_i^\dagger c_i^{\phantom\dagger}$ is the itinerant electron number operator and $n_{if}$ is the localized electron number operator; we work in the infinite-dimensional limit, where $D\rightarrow \infty$ with $t^*$ remaining finite. The hopping is between nearest neighbor sites $i$ and $j$.
We assume that the electric field is spatially uniform, pointing along the diagonal, with $A(r,t)=A(t)$ the component in each spatial direction, and we ignore magnetic field and relativistic effects. The vector potential in the Hamiltonian gauge then produces the total electric field through $\mathbf{E}_{\text{tot}}(t)=-d\mathbf{A}(t)/dt$ (we set $\hbar=c=e=a=1$). 

This vector potential has two terms $\mathbf{A}(t)=\mathbf{A}_{\text{pump}}(t)+\mathbf{A}_{\text{probe}}(t)$. 
We model the pump field by
\begin{equation}\label{efield}
\mathbf{E}_\text{{pump}}(t)=\mathbf{E}_0\cos{[\omega_p (t-t_p)]} e^{-\frac{(t-t_p)^2}{\sigma_p^2}},
\end{equation}
where $\mathbf{E}_0$ is the magnitude of the pump field at time $t=t_p$, and $\omega_p$ and $\sigma_p$ define the frequency and width of the pump pulse, respectively.   

Electronic Raman scattering is a second-order process, so we find resonant and nonresonant terms that are second order in
the probe vector potential $\mathbf{A}_{\text{probe}}(t)$~\cite{shastry:1068,freericks:125110,shvaika:137402,shvaika:045120,devereaux:175}. We will only work with the nonresonant contribution here. Summing over all the quantum states between the start and the end of the experiment, we end up with the final state in Eq.~(\ref{evolution}) 
\begin{align}\label{final_state}
& |\psi(\infty)\rangle = \frac{1}{2}\int^{\infty}_{-\infty}d\tilde{t}U_{\text{pump}}(t_{\text{exp}},\tilde{t})A^{\alpha}_{\text{probe}}(\tilde{t})\nonumber \\
 &~~\times\gamma_{\alpha\beta}(\tilde{t})A^{\beta}_{\text{probe}}(\tilde{t})
  U_{\text{pump}}(\tilde{t},-\infty) |n\rangle\otimes a_{\mathbf{k}_i, \mathbf{e}_i}^\dagger|0\rangle,
\end{align}
where the stress tensor operator in the momentum representation is $\gamma_{\alpha\beta}(t)=\sum_{k}\frac{\partial^{2} \epsilon(\mathbf{k},t)}{\partial k_{\alpha}\partial k_{\beta}}c_{\mathbf{k}}^{\dag}c_{\mathbf{k}}^{\phantom{\dagger}}$, with a time-dependent electronic band energy $\epsilon(\mathbf{k},t)=-\frac{t^*}{\sqrt{D}}\sum_{\alpha=1}^{D}\cos{[k_{\alpha}-A^{\alpha}_{\text{pump}}(t)]}$ (the repeated indices $\alpha$ and $\beta$ are summed over). The probe field $A_{\text{probe}}^{\alpha}(t)=s(t)\sum_{\mathbf{k}, \mathbf{e}} (\frac{2\pi}{\omega_{\mathbf{k}}})^{1/2} e_{\alpha}(e^{i\omega_{\mathbf{k}} t} a_{\mathbf{k},\mathbf{e}}^{\dag}+e^{-i\omega_{\mathbf{k}} t} a_{\mathbf{k}, \mathbf{e}})^{\phantom{\dagger}}$, which acts in the photon space,  describes the creation and annihilation of photons with polarization $\mathbf{e}$, and frequency $\omega_{\mathbf{k}}$. The time profile of the probe pulse is defined by an envelope function $s(t)$, which we take to be $s(t)=\exp{[-(t-t_0)^2/\sigma^{2}_{\text{b}}]}/(\sigma_{\text{b}}\sqrt{\pi})$ centered on time $t_0$ (which defines time delay of the probe). The width of the probe pulse is $\sigma_{\text{b}}$. 
The operator $U_{\text{pump}}(t_2,t_1)$ is the evolution operator in Eq.~(\ref{evolution}), but without the probe pulse.

The scattering amplitude is defined by the probability to find a photon with energy $\omega_f$ and polarization $\mathbf{e}_f$ in the  final state given in Eq.~(\ref{final_state}). This is then weighted by the thermal factors and summed over all the equilibrium states. The electronic Raman scattering probability becomes
\begin{align}\label{raman_probability}
R(\omega_i-\omega_f,t_0)=\sum\limits_{n}\frac{e^{-\beta E_n}}{\mathcal{Z}}
 \langle \psi(\infty)|a_{\mathbf{k}_f,\mathbf{e}_f}^{\dag}a_{\mathbf{k}_f,\mathbf{e}_f}^{\phantom{\dagger}}|\psi(\infty)\rangle, 
\end{align}
where $\mathcal{Z}=\Tr \exp{(-\beta\mathcal{H}_0)}$ is the partition function at the initial temperature $T=1/\beta$,
and one needs to calculate an expectation value for the scattering probability over the photon vacuum state $|0\rangle$. 
Applying the Kadanoff-Baym-Keldysh formalism~\cite{kadanoff:1962,keldysh:1018} we introduce Green's functions which are built on two stress tensor operators $R^{c}_{\gamma\gamma}(t,t')=-i\Tr \exp{(-\beta\mathcal{H}_0)}\mathcal{T}_c\gamma(t)\gamma(t')/\mathcal{Z}$ with times $t$ and $t'$ being ordered on the contour. In the same way as it was done by Nozi\`{e}res and Abrahams~\cite{nozieres:3099} for RIXS, one can show that this greater Green's function defines the electronic Raman scattering probability $R(\Omega,t_0)$ when the two  times $t$ and $t'$ are placed on different branches of the contour with $t$ ahead of $t'$:
\begin{equation}\label{raman_greater}
R^{>}_{\gamma\gamma}(t,t')=-i\frac{1}{\mathcal{Z}}\Tr e^{-\beta\mathcal{H}_0}\gamma(t)\gamma(t'). 
\end{equation}
Diagrammatically, the greater function $R^{>}_{\gamma\gamma}(t,t')$ 
consists of the bare and renormalized bubbles, which are constructed from the greater $G^{>}_k(t,t')=-i\Tr \exp{(-\beta\mathcal{H}_0)} c_k^{\phantom{\dag}}(t)c_k^{\dag}(t')/\mathcal{Z}$ and the lesser $G^{<}_k(t,t')=i\Tr \exp{(-\beta\mathcal{H}_0)} c_k^{\dag}(t')c_k^{\phantom{\dag}}(t)/\mathcal{Z}$ momentum-dependent single-particle Green's functions. The vertices of the bubbles
include the factor $\bar{\gamma}(\mathbf{k},t)=\sum_{\alpha\beta}e_{i\alpha}\frac{\partial^{2}\epsilon(\mathbf{k}-\mathbf{A}_{\text{pump}}(t))}{\partial k_{\alpha}\partial k_{\beta}}e_{f\beta}$. 
We consider $B_{\text{1g}}$ symmetry only, so the polarization vectors of the incident and scattered photons are equal to $\mathbf{e}_i=(1,1,1,...)$ and $\mathbf{e}_f=(-1,1,-1,...)$, respectively. In the case of nearest neighbor hopping, we find $\bar{\gamma}_{B_{\text{1g}}}(\mathbf{k},t)=\sum_{\alpha=1}^{D}(-1)^{\alpha}\cos{[k_{\alpha}-A^{\alpha}_{\text{pump}}(t)]}/{\sqrt{D}}$. Due to this form of the stress tensor and the local character of the irreducible charge vertex~\cite{zlatic:263}, the renormalized bubble vanishes~\cite{freericks:149,shvaika:045120} 
and we end up with the bare bubble only.
 
We perform the summation over momenta $\mathbf{k}$ by integrating over energy with the joint density of states $\rho(\epsilon)\rho(\bar\epsilon)$, which is the product of the Gaussians given by $\rho(\epsilon)= \exp{(-\epsilon^2)}/\sqrt{\pi}$ and $\bar\epsilon=\bar\epsilon(\mathbf{k})=-\sum_{\alpha=1}^{D}\sin k_{\alpha}/\sqrt{D}$~\cite{freericks:149,freericks:125110}.
Finally, we perform the Fourier transform from time to frequency and we obtain [after suppressing the overall ``scattering strength'' prefactor $4\pi^2/(\omega_i\omega_f)$]:
\begin{widetext}
\begin{equation}\label{ram_cs}
R_{B_{\text{1g}}}^{N}(\Omega,t_0)=\frac{1}{2}\Real\int \!d t\! \int \!d t' s^2(t) s^2(t') e^{i\Omega(t-t')}\cos{[A_{\text{pump}}(t)-A_{\text{pump}}(t')]} \int \!d\epsilon\!\int \!d\bar\epsilon \rho(\epsilon)\rho(\bar\epsilon)G_{\epsilon,\bar\epsilon}^{>}(t,t')G_{\epsilon,\bar\epsilon}^{<}(t',t) 
\end{equation}
\end{widetext}
with $\Omega=\omega_i-\omega_f$ being the frequency shift for the scattered photons. Note that this is a general DMFT result, independent of the choice of Hamiltonian.

In equilibrium and for the monochromatic light beams ($s(t)\to$ constant), the Stokes and anti-Stokes lines of the Raman cross section are connected by the relation $R_{B_{\text{1g}}}^{N}(\Omega)/R_{B_{\text{1g}}}^{N}(-\Omega)= e^{\beta\Omega}$, for $\Omega > 0$. When a probe pulse is present,  this relation is replaced by a similar one with a displaced frequency $\Tilde\Omega = \Omega - \frac{\beta}{\sigma_b^2}$~\cite{shvaika:33707}
\begin{equation}\label{stokes-antistokes}
\frac{R_{B_{\text{1g}}}^{N}(\Tilde\Omega-\frac{\beta}{\sigma_b^2},t_0)}{R_{B_{\text{1g}}}^{N}(-\Tilde\Omega-\frac{\beta}{\sigma_b^2},t_0)}= e^{\beta\Tilde\Omega}. 
\end{equation}
We have found from our numerics, that in a wide enough region of $\Tilde\Omega \approx 0$, the value of the ratio in Eq.~(\ref{stokes-antistokes}) holds and can be used to estimate the effective temperature $\beta_{\text{eff}}=1/T_{\text{eff}}$ of the two-particle excitations during the nonequilibrium process.
Motivated by~\cite{yang:40876}, we can compare this ``two-particle'' temperature, with an effective ``single-particle'' temperature extracted from the time-resolved photoemission spectra (tr-PES). The tr-PES spectral function is defined by the local lesser Green's function as follows~\cite{freericks:136401}:
\begin{equation}\label{pes_les}
 P^<(\omega,t_0)=-i\int \!dt\!\int \!dt' s(t) s(t') e^{-i\omega(t-t')}G^{<}_{\text{loc}}(t,t');
\end{equation}
the local Green's function is found from the momentum-dependent Green's function by summing over all momenta with equal weight.
As an analogy to the equilibrium case, 
we define the nonequilibrium density-of-states via the probe-envelope-modified retarded response function:
\begin{align}\label{pes_ret}
 A_d(\omega,t_0)&=i\int dt\int dt' s(t) s(t') e^{-i\omega(t-t')}\nonumber \\
 &\times\theta(t-t')[G^{>}_{\text{loc}}(t,t')-G^{<}_{\text{loc}}(t,t')]. 
\end{align}
The ratio of the tr-PES spectral function to density-of-states is used to determine the nonequilibrium distribution function for the fermionic states~\cite{smallwood:235107,stafford:245403} $f_s(\omega,t_0)=P^<(\omega,t_0)/A_d(\omega,t_0)$.

\begin{figure*}[!htb]
	\includegraphics[width=0.29\textwidth]{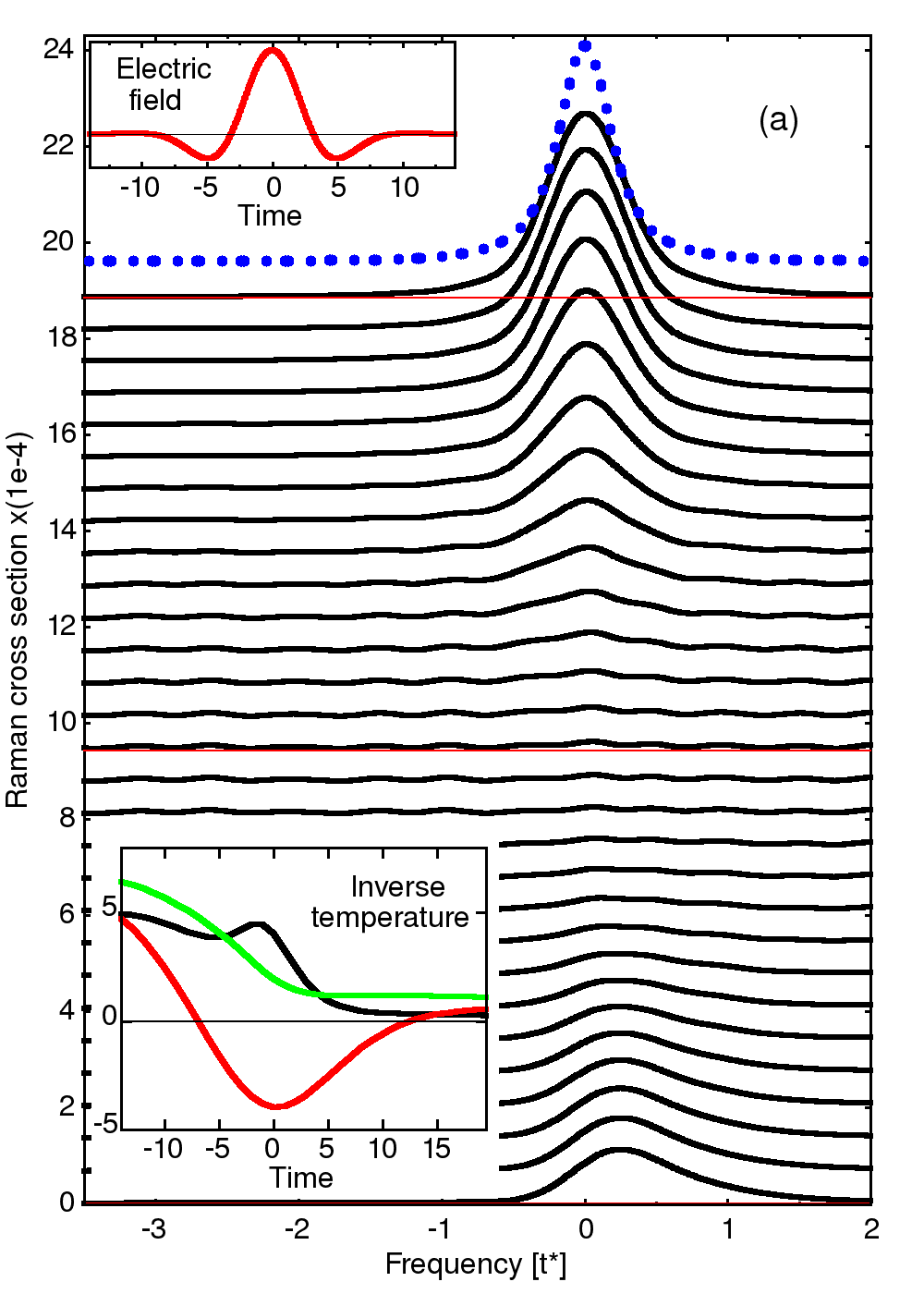}\hspace{-0.3cm}
	\includegraphics[width=0.29\textwidth]{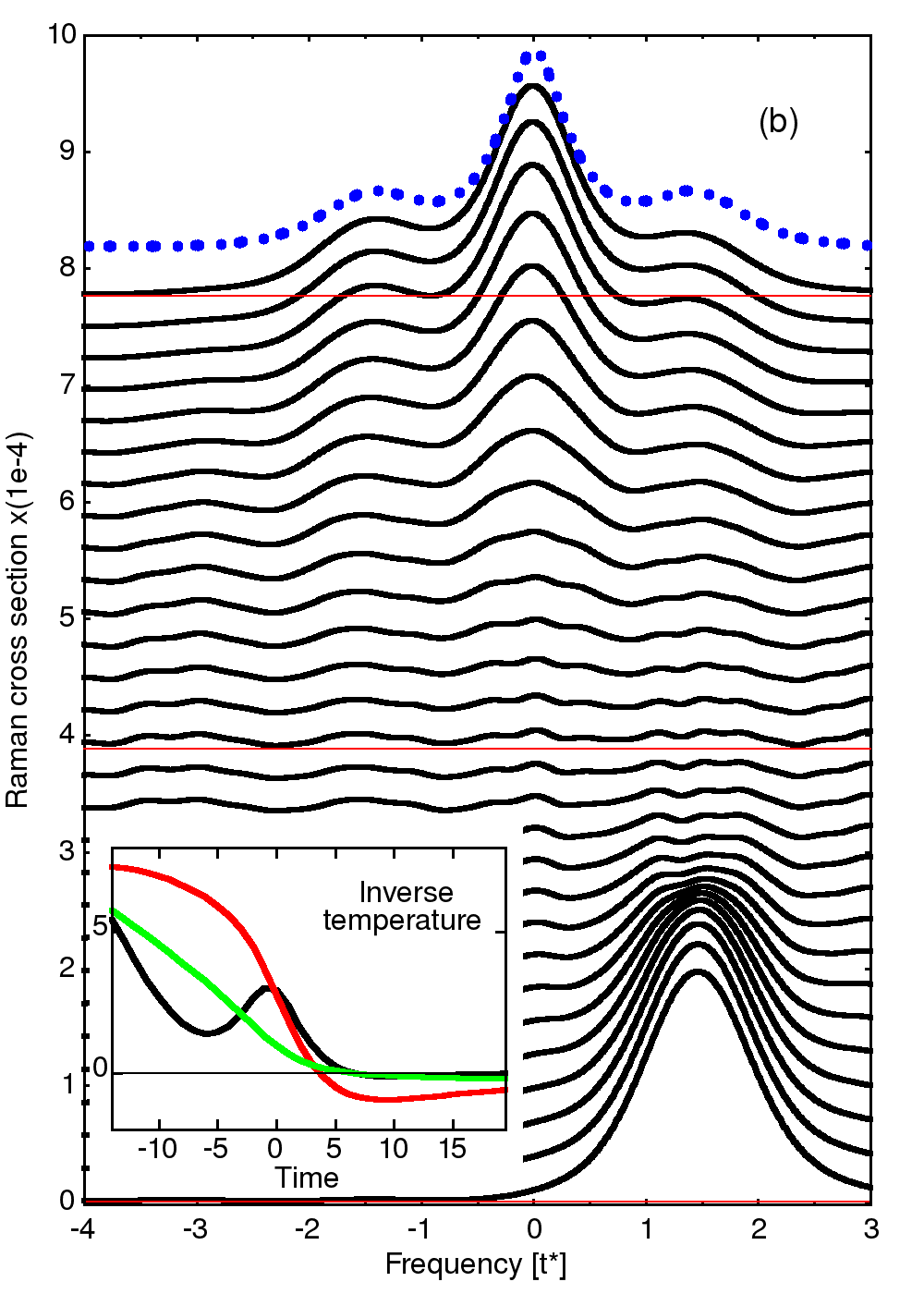}\hspace{-0.3cm}
	\includegraphics[width=0.29\textwidth]{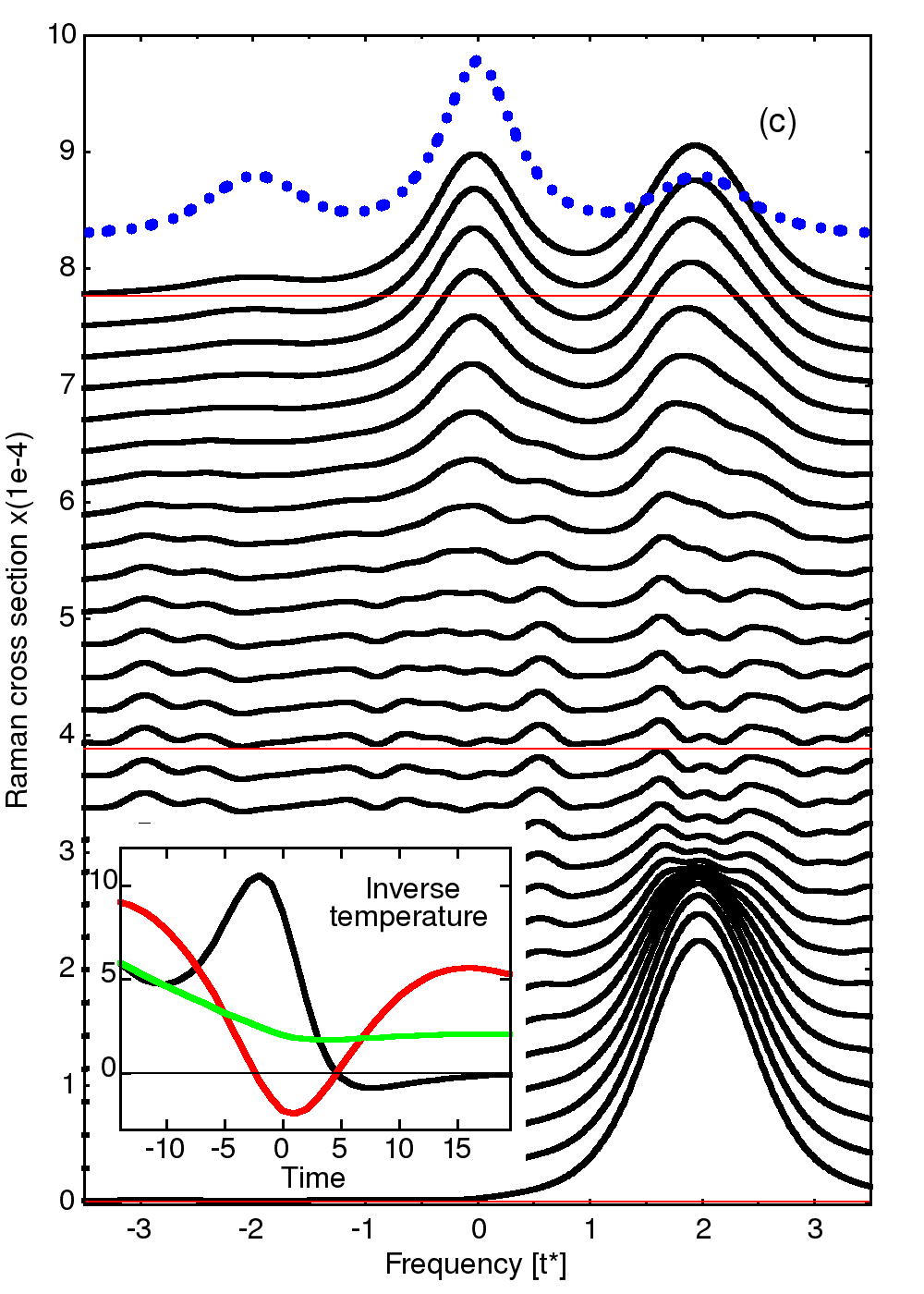}
	\caption{Nonresonant Raman scattering cross section for (a) $U=0.5$, (b) $U=1.5$, and (c) $U=2.0$. Different lines correspond to different time delays $t_0\in[-14,14]$ of the probe pulse with respect to the pump one. The pump pulse field is plotted in the upper inset in panel (a) with the following parameters: $\mathbf{E}_0=30$, $t_p=0$; $\omega_p=0.5$, and $\sigma_p=5$. The probe pulse width is $\sigma_b=12$. The infinite-temperature limit is shown with the blue dotted lines. The lower insets shows the time dependence of the inverse temperature: the black curve is from the Stokes/anti-Stokes ratio; the red curve is from the slope of the single-particle distribution function and the green curve is from a least squares fit to the Fermi-Dirac distribution function; the initial temperature is $T=0.1$ ($\beta=1/T=10$), but the effective temperature only approaches that value for earlier experimental times~\cite{shvaika:33707}.}
	\label{raman}
\end{figure*}

We present our results for the nonresonant Raman scattering cross section in Fig.~\ref{raman}: panel (a) is a metal, panel (b) is a near critical Mott-insulator, and panel (c) is a Mott-insulator.  For all three cases, we see some common behavior; in equilibrium, the electronic Raman scattering has a broad peak set by the bandwidth for the metal, which evolves into a Mott peak centered at $U$ for the Mott insulator. At early times, there is one peak near $\Omega=U$ (Stokes line) which corresponds to the charge-transfer peak in equilibrium. As the pump is applied $(t_0\approx 0)$, it completely suppresses this process due to strong Bloch oscillations of the stress tensor $\gamma_{\alpha\beta}(t)$~\cite{shvaika:33707}. After the pump is gone the Raman scattering becomes different for different $U$'s. At late times there is always a peak around $\Omega=0$,  
which is no longer Pauli blocked due to the excitation of electrons from the pump and comes from hot electrons generated by the pump pulse. 

In the insets, we present the time evolution of the effective inverse temperatures $\beta_{\text{eff}}=1/T_{\text{eff}}$ for the single-particle (red and green lines) and two-particle (black line) excitations and one can see that at late times $\beta_{\text{eff}}$ becomes very small. We also plot the $T\to\infty$ equilibrium Raman cross section with the blue dotted line, which is close to the nonequilibrium one at late times. For large $U=2.0$ (Fig.~\ref{raman}(c)) there are two more peaks after the pump is gone: the charge-transfer Stokes and anti-Stokes peaks at $\Omega=\pm U$. In the near-critical Mott-insulator with $U=1.5$ in Fig.~\ref{raman}(b), there are three peaks at late times that correspond to the ``zero-peak'' and to the restored charge-transfer peaks at $\Omega=U$ (Stokes) and at $\Omega=-U$ (anti-Stokes). Moreover, the anti-Stokes' peak is higher than the Stokes' peak, which implies a negative temperature.       

\begin{figure*}[!htb]
	\includegraphics[width=0.29\textwidth]{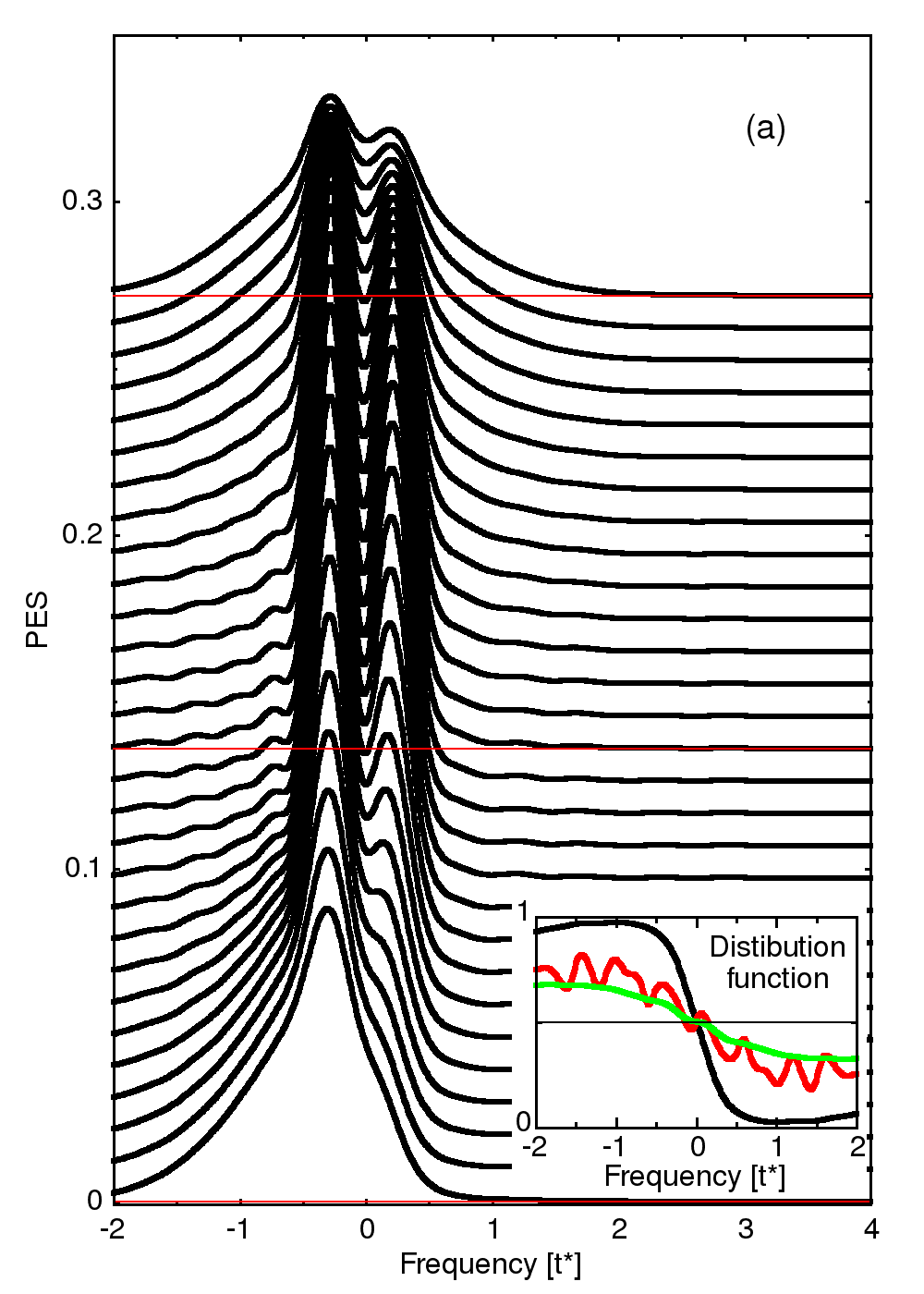}\hspace{-0.3cm}
	\includegraphics[width=0.29\textwidth]{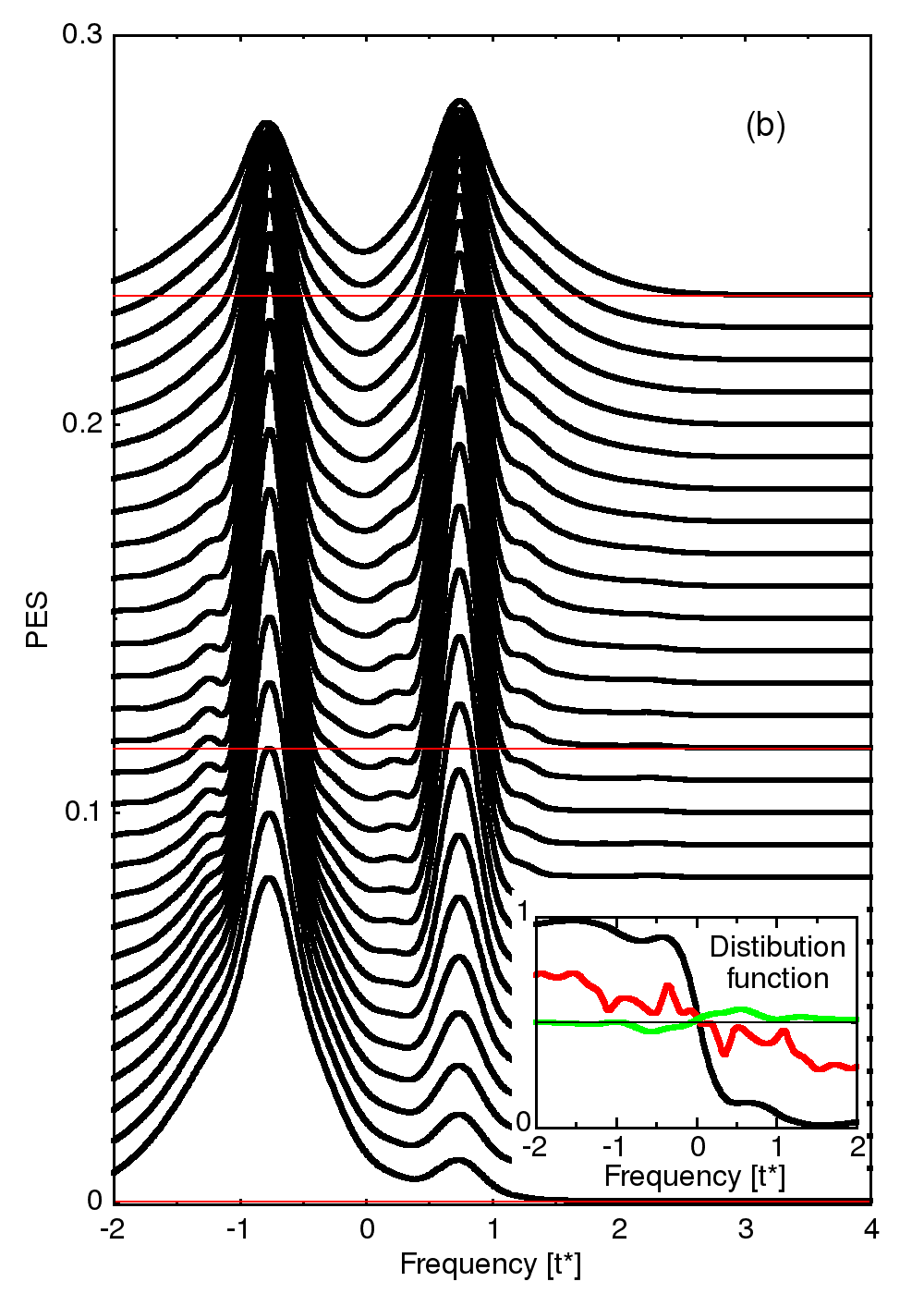}\hspace{-0.3cm}
	\includegraphics[width=0.29\textwidth]{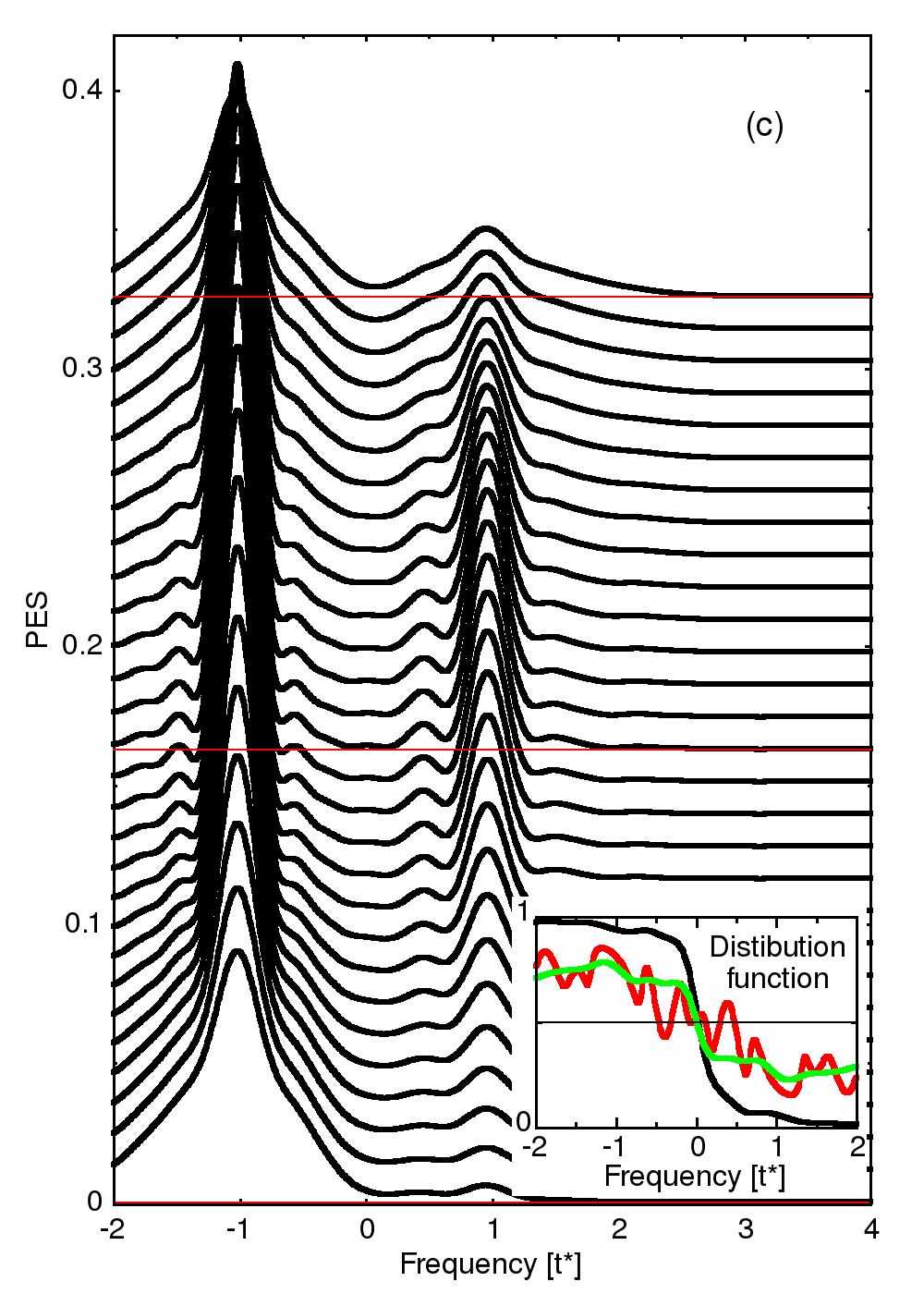}
	\caption{tr-PES: (a), (b), and (c) correspond to $U=0.5$, $U=1.5$, and $U=2.0$. Different lines correspond to different time delays $t_0\in[-14,14]$. 
	The inset shows the fermionic distribution function at $t_0=-14$ (black), $t_0=0$ (red), and $t_0=14$ (green).}
	\label{pes}
\end{figure*}

For comparison, we present our results for the tr-PES in Fig.~\ref{pes}. Similar to the Raman scattering in Fig.~\ref{raman}, the three different panels correspond to different $U$ values, and different curves correspond to different delay times $t_0$. In the insets, we show the nonequilibrium distribution function at times $t_0=-14$ (black curve), $t_0=0$ (red curve), and $t_0=14$ (green curve). 
The pump field excites electrons to the upper band, and after the pump is gone they de-excite back to the lower band in the cases of $U=0.5$ and $U=2.0$ in Fig.~\ref{pes} (a) and (c), respectively. But in the case of the near-critical Mott-insulator at $U=1.5$ in Fig.~\ref{pes} (b) it is the opposite: the magnitude of the tr-PES response from the upper band is larger than from the lower one at late times. This means that we observe an inverse occupation of the single-particle electron states characterized by a negative temperature. This is also seen from the plots for the nonequilibrium distribution function: the slope of the green curve which corresponds to late time $t=14$ at zero frequency in Fig.~\ref{pes} (b) has an opposite sign with respect to other curves on this inset as well as to those in panels (a) and (c). 

In the bottom insets in Fig.~\ref{raman}, we show the effective inverse temperature calculated from the Stokes to anti-Stokes ratio in Eq.~(\ref{stokes-antistokes}) (black curve) 
and from the slope of the nonequilibrium distribution function $f_s(\omega, t_0)$ at the Fermi level, which we extracted from the tr-PES results using Eqs.~(\ref{pes_les}) and (\ref{pes_ret}) (red curve) and by least squares interpolation of the Fermi-Dirac distribution (green curve). Exploring the behavior of the inverse temperature during the pump, one may speculate on the sensitivity of the single-particle excitations measured by tr-PES and the two-particle excitations measured by Raman scattering to nonequilibrium pumping. The population of the single-particle states is changed very rapidly with the pump for small values of the Coulomb interaction $U=0.5$ and changes very slowly for the large gap Mott insulator at $U=2$, whereas for the near-critical Mott insulator $U=1.5$ the gap is small enough to allow fast population of the upper band but large enough to prevent back de-excitation, leading to an inverse occupation and negative temperatures. At later times the effective temperatures are high leading to the creation of hot electrons as manifested by the flattened electronic distribution function and by the central peak in the Raman cross section. The two-particle excitations are a signature of the creation of bound states, which behave like a heavy subsystem with a different effective temperature and relaxation time than the single-particle one. 

Note how the effective temperature of the two-particle excitations initially increases with the pump, but then starts to decrease, reaching its minimum at the pump maximum, after which it starts to increase again. Such behavior can be explained by two effects. First, it is the consequence of a suppression of Raman scattering by Bloch oscillations and, second, the hot electrons destroy the bound states decreasing their density which, together with rapid heating of light single-particle excitations, leads to adiabatic cooling of the heavy (two-particle) subsystem. Of course, because the isolated Falicov-Kimball model does not thermalize, the green and black curves never agree at long times, but they can become close when one is near infinite temperature.

In conclusion, we employed the theory for nonresonant electronic Raman scattering in the $B_{\text{1g}}$ symmetry channel to show how one can measure both fermionic and collective bosonic temperatures stroboscopically in a pump/probe experiment. By comparing these effective temperatures to each other, we can determine how far from equilibrium the electrons are (since these two temperatures must agree in equilibrium). Given the fact that a good fit to a Fermi-Dirac distribution may still involve nonequilibrium electrons, this consistency test across fermions and collective bosonic excitations provides a stringent test for the approach to thermal equilibrium.

\textit{Acknowledgments}:
This work was supported by the Department of Energy, Office of Basic Energy Sciences, Division of Materials Sciences and Engineering under Contract Nos. DE-AC02-76SF00515 (Stanford/SIMES) and DE-FG02-08ER46542 (Georgetown). Computational resources were provided by the National Energy Research Scientific Computing Center supported by the Department of Energy, Office of Science, under Contract No. DE- AC02-05CH11231. J.K.F. was also supported by the McDevitt bequest at Georgetown.

\bibliography{raman_ref}

\end{document}